\documentclass[preprints,article,accept,moreauthors,pdftex]{mdpi}
\firstpage{1} 
\pubvolume{1}
\issuenum{1}
\articlenumber{0}
\pubyear{2023}
\copyrightyear{2023}
\datereceived{} 
\dateaccepted{} 
\datepublished{} 
\hreflink{https://doi.org/} % If needed use \linebreak
\Title{A Community Detection and Graph Neural Network Based Link Prediction Approach for Scientific Literature}

\TitleCitation{A Community Detection and Graph Neural Network Based Link Prediction Approach for Scientific Literature}

\Author{Chunjiang Liu $^{1, \ddagger,}$*\orcidA{}, Yikun Han $^{2, \ddagger}$\orcidB{}, Haiyun Xu $^{3}$\orcidC{}, Shihan Yang $^{4}$\orcidD{} ,  Kaidi Wang $^{5}$\orcidE{} and Yongye Su $^{6}$\orcidF{}}

\AuthorNames{Firstname Lastname, Firstname Lastname , Firstname Lastname , Firstname Lastname , Firstname Lastname and Firstname Lastname}

\AuthorCitation{Liu, C.; Han, Y.; Xu, H.; Yang, S.; Wang, K.; Su, Y.}
\address{%
$^{1}$ \quad Chengdu Library and Information Center, Chinese Academy of Sciences, Chengdu 610299, China; liucj@clas.ac.cn\\
$^{2}$ \quad Department of Statistics, University of Michigan, Michigan 48109, United States; yikunhan@umich.edu\\
$^{3}$ \quad School of Business, Shandong University of Technology, Zibo 255000, China; xuhaiyunnemo@gmail.com\\
$^{4}$ \quad Faculty of Management and Economics, Kunming University of Science and Technology, Kunming 650031, China; dr.yangsh@kust.edu.cn\\
$^{5}$ \quad School of Business, Macau University of Science and Technology, Macau 999078, China; kdwang@must.edu.mo\\
$^{6}$ \quad Department of Computer Science, Purdue University, Indiana 47907, United States; su311@purdue.edu}

\corres{Correspondence: liucj@clas.ac.cn}
\secondnote{These authors contributed equally to this work.}
\abstract{This study presents a novel approach that synergizes community detection algorithms with various Graph Neural Network (GNN) models to bolster link prediction in scientific literature networks. By integrating the Louvain community detection algorithm into our GNN frameworks, we consistently enhance performance across all models tested. For example, integrating Louvain with the GAT model resulted in an AUC score increase from 0.777 to 0.823, exemplifying the typical improvements observed. Similar gains are noted when Louvain is paired with other GNN architectures, confirming the robustness and effectiveness of incorporating community-level insights. This consistent uplift in performance—reflected in our extensive experimentation on bipartite graphs of scientific collaborations and citations—highlights the synergistic potential of combining community detection with GNNs to overcome common link prediction challenges such as scalability and resolution limits. Our findings advocate for the integration of community structures as a significant step forward in the predictive accuracy of network science models, offering a comprehensive understanding of scientific collaboration patterns through the lens of advanced machine learning techniques.}

\keyword{link prediction; graph neural network; community detection; network analysis; citation graph; deep learning} 

\begin{document}
% \tableofcontents
%%%%%%%%%%%%%%%%%%%%%%%%%%%%%%%%%%%%%%%%%%
% \setcounter{section}{-1} %% Remove this when starting to work on the template.
% \section{How to Use this Template}

% The template details the sections that can be used in a manuscript. Note that the order and names of article sections may differ from the requirements of the journal (e.g., the positioning of the Materials and Methods section). Please check the instructions on the authors' page of the journal to verify the correct order and names. For any questions, please contact the editorial office of the journal or support@mdpi.com. For LaTeX-related questions please contact latex@mdpi.com.%\endnote{This is an endnote.} % To use endnotes, please un-comment \printendnotes below (before References). Only journal Laws uses \footnote.

% The order of the section titles is different for some journals. Please refer to the "Instructions for Authors” on the journal homepage.

\section{Introduction}

Graphs represent intricate data structures comprised of vertices and edges, where link prediction aims to foresee potential connections and their strengths within a graph's structure \cite{Liben2007LinkPrediction}. While traditional methods like Markov chains and machine learning have found utility in applications such as website browsing and citation prediction \cite{Sarukkai2000MarkovChains} \cite{Popescul2003StatisticalRelational}, they often rely heavily on node attributes, which can be challenging to obtain and validate in real-world settings. Consequently, these methods may not always yield high accuracy in predictions. Graph embedding techniques, by contrast, can effectively encapsulate complex structures and relational dynamics, paving the way for more accurate applications of link prediction algorithms, now an active area of research.

In this paper, we introduce a pioneering approach that unifies community detection algorithms with Graph Neural Networks (GNN) to refine link prediction within scientific literature networks. Employing the Louvain algorithm, we reveal latent community structures that GNN models can exploit to enhance link prediction precision. This methodology is particularly potent in scientific collaboration networks, where understanding and forecasting collaborative patterns is crucial. For instance, our approach could significantly improve paper recommendation systems, aiding researchers in finding relevant studies and potential citations when preparing their scientific manuscripts \cite{farber2020citation}. Such systems are invaluable given the recent exponential growth in scientific publications and the critical role of citing pertinent work.

\begin{figure}[H]
\centering
\includegraphics[width=13.5cm]{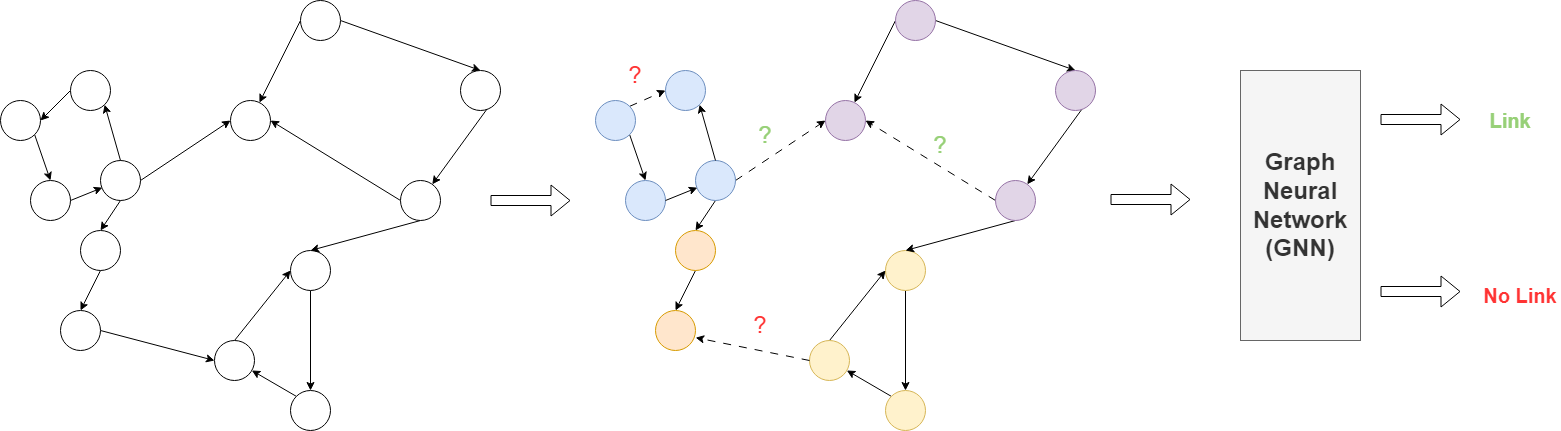}
\caption{General idea which integrates community detection algorithms with GNN models.\label{idea}}
\end{figure}

Our integrated approach, succinctly depicted in Figure \ref{idea}, marks a substantial advancement in predictive network analysis. It is especially pertinent to the scientific literature domain, where discerning and anticipating collaboration and citation dynamics can offer profound insights, potentially fostering novel research intersections and driving the progress of scientific inquiry forward.

The structure of the paper is as follows: Section 2 discusses related work to frame the current advancements in link prediction and graph neural networks. Section 3 explains the datasets, data collection, and preprocessing steps. Section 4 delves into our methods, firstly examining heuristic-based link prediction techniques, then exploring machine learning methods, followed by a detailed look at our advanced graph neural network approaches with community detection. Section 5 discusses the broader implications, limitations, and avenues for future research arising from our study. The paper concludes in section 6, summarizing our contributions and reflecting on the potential impact on network analysis in the realm of scientific literature.
%%%%%%%%%%%%%%%%%%%%%%%%%%%%%%%%%%%%%%%%%%
\section{Related Work}
\subsection{Graph Neural Network}

The graph structure in computers belongs to a non-Euclidean structure, which is characterized by irregular arrangement, and it is difficult to define its neighboring nodes for a certain point in the data, and the number of neighboring nodes of different nodes is also different, so it is difficult to be directly applied to deep learning methods such as convolutional neural networks and recurrent data networks. The graph embedding model aims to downscale the graph structure, i.e., mapping a highly dense graph into a low-density graph. Graph embedding algorithms are now widely used in the fields of node classification, link prediction, clustering and visualization \cite{Bhagat2011NodeClassification}.

Due to the different graph structures and different algorithmic principles applied in the embedding process, graph embedding models are mainly categorized as follows. Cai et al. classified graph embedding models into four categories according to the different structures of the input graphs, i.e., matrix decomposition-based graph embedding, deep learning-based graph embedding, edge reconstruction-based graph embedding, graph kernel-based graph embedding, and generative model-based graph embedding \cite{Cai2017GraphEmbeddingSurvey}. Goyal et al. classified graph embedding into three categories according to the model utilized in the embedding process, i.e., factorization-based graph embedding model, random walk-based graph embedding model, and deep learning-based graph embedding model \cite{Goyal2017GraphEmbedding}. Chen et al. classified graph embedding models into seven categories by incorporating the type of the graph along with the model utilized in the embedding process as the basis for the classification, i.e., dimensionality reduction-based graph embedding, random walk-based graph embedding, matrix decomposition-based graph embedding, neural network-based graph embedding, graph embedding for large graphs, graph embedding for hypergraphs, and graph embedding based on attentional mechanisms \cite{Chen2020GraphRepresentation}. Xie et al. classified the dynamic networks into five categories based on whether they are of snapshot type or continuous time type, and in accordance with their different graph embedding strategies, i.e.: eigenvalue decomposition-based graph embedding, Skim-Gram model-based graph embedding, self-encoding-based graph embedding, and graph convolution-based graph embedding \cite{Xie2020DynamicNetwork}. Xia et al. classified graph embedding algorithms into four categories based on graph learning techniques, i.e., graph embedding for graph signal processing, graph embedding for matrix decomposition, graph embedding for random walk, and graph embedding for deep learning \cite{Xia2021GraphLearning}. Based on the existing research ignoring the correlation between static and dynamic graphs, Yuan et al. simultaneously classified static and dynamic graph models into five categories based on the algorithmic principles applied in the embedding process, namely: graph embedding based on matrix decomposition, graph embedding based on random walk, graph embedding based on self-encoder, graph embedding based on graph neural network, and graph embedding based on other methods \cite{Yuan2022GraphEmbeddingModel}.

The following summarizes the current mainstream graph embedding algorithm models at home and abroad, and briefly introduces their principles. Singular value decomposition (SVD) is a typical algorithm for the decomposition of adjacency matrices, and SVD essentially employs the idea of vector orthogonal transformation to decompose matrices. Further, on the basis of generalized SVD (GSVD), combined with asymmetric transfer, the HOPE algorithm is derived \cite{Ou2016AsymmetricTransitivity}. Node2vec \cite{Grover2016Node2vec}, as a classical algorithm for random walk, is inspired by the skip-gram algorithm and DeepWalk algorithm in the language model, and artificially introduces two parameters on the basis of obtaining sequences of neighboring nodes via random walk along the balance the breadth-first traversal and depth-first traversal, so as to take into account the homogeneity and structure of the graph. Due to the strong time-varying nature of the network in real problems, the CTDNE algorithm \cite{Nguyen2018DynamicNetwork} introduces temporal information on the basis of static graph random walk, i.e., random walk based on the temporal path, which is more suitable for dynamic graph scenarios. Based on the feature decomposition of the Laplace matrix, GCN \cite{Kipf2016GraphConvolutional} simultaneously deals with the node features as well as the connections between nodes to realize a new node representation. Based on GCN, GraphSAGE \cite{Hamilton2017InductiveLearning} utilizes known nodes to predict unknown nodes through a fixed-size sampling mechanism, and GAT \cite{Velikovic2017GraphAttention} introduces an attention mechanism to assign weights to nodes so that the input of complete graph data is not required, which is also more in line with real-world scenarios.

In addition, since many real-world problems are not pairwise node relations, concepts such as hypergraphs and heterogeneous graphs have arisen. In order to better study these graph structures, some scholars have also proposed graph embedding algorithms for hypergraphs, e.g., DHNE \cite{Ke2017StructuralDeepEmbedding}, HGNN \cite{Feng2018HypergraphNN}, etc., as well as algorithms for heterogeneous graphs, e.g., HGSL \cite{Tang2023Hypergraphs}, etc.

Building on the extensive survey of graph embedding techniques presented in Section 2, our study specifically focuses on the advancements of GNN methods, particularly GAT, GCN, and GraphSAGE. These methods represent the most directly related prior work to our study due to their state-of-the-art performance in various graph-based tasks and their relevance to our proposed approach. In the context of these foundational GNN models, our work introduces a novel enhancement by integrating the Louvain method for community detection. This integration significantly boosts the models' performance across multiple metrics, including precision, recall, F1 score, and AUC score. The Louvain method's ability to uncover latent community structures provides an additional layer of information, enriching the original node features and thereby enabling the GNN models to achieve a deeper understanding of the graph's topology. This enhancement aligns with the recent trends in deep learning that prioritize the incorporation of structural and contextual information within the learning process, affirming our contribution to the evolving landscape of graph neural network research.

\subsection{Link Prediction}
The following key metrics are mainly used in link prediction algorithms, i.e., Area Under the Curve (AUC) metric, Accuracy, Common Neighbor (CN) metric, Adamic-Adar metric, Resource Allocation metric (RA), Preferred Attachment metric (PA), Random Walk with Restart (RWR) metric and Katz metric. Traditional link prediction algorithms can be divided into two categories: unsupervised and supervised algorithms, where unsupervised algorithms are mainly centered around generating functions, node clustering, likelihood analysis, and other related models, while models under supervised algorithms are more diverse. In recent years, many correlations in graph embedding algorithms have been utilized in link prediction, e.g., DeepWalk algorithm, LINE algorithm. In solving practical problems, since the network structure tends to be more complex and diverse, the main research directions of link prediction models are currently focused on temporal network link prediction, heterogeneous network link prediction and directed graph link prediction.

Links in temporal networks change over time and thus have a more complex and diverse dynamic structure. For link prediction in time-series networks, Rahman et al. transformed the link prediction problem into an optimal coding problem for graph embedding through feature representation extraction, and minimized the reconstruction error through gradient descent, and proposed the Dylink2vec algorithm. This method simultaneously considers the topology of the network and the historical changes of the links, and compares to the traditional link prediction methods based on a single topological structure and DeepWalk based link prediction methods, Dylink2vec experiments yielded higher CN, AA and Katz metrics \cite{Rahman2018Dylink2vec}. Lekui Zhou et al. proposed the DynamicTriad algorithm in conjunction with representational learning, which can simultaneously preserve the network's structural and dynamic change information, and through the triadic closure process, the nodes disconnected from the triad are connected to each other, in order to capture the dynamic changes at each moment and performs representation learning \cite{Zhou2018DynamicNetwork}. Wu et al. established the GETRWR algorithm, which draws on the random walk algorithm in graph embedding, and improves the transfer probability in the RWR metrics by applying the node2vec model, and by introducing a restart factor, i.e., the particles have the probability to return to the initial position in the process of random walk and dynamically calculate the transfer probability of the particles to achieve higher indicators such as AUC and RWR \cite{Wu2020LinkPrediction}.

Nodes and edges in heterogeneous networks have different data types, and many link predictions based on homogeneous networks are often not directly applicable. For link prediction in heterogeneous networks, Dong et al. proposed the metapath2vec algorithm, which firstly uses a random walk model based on metapaths to construct neighboring nodes in the network, and then performs graph embedding operations on them using the skip-gram algorithm, and finally predicts the neighboring nodes of the nodes through a method based on negative sampling of heterogeneous networks. This algorithm can maximize the preservation of the structure and semantics of the heterogeneous network \cite{Dong2017Metapath2vec}. Fu et al. on the other hand, through the construction of the neural network model used for heterogeneous network representation learning, and then use the Hin2vec algorithm to capture the different node relationships and network structure in it, combined with the meta-path to the nodes, and then the vector feature learning to generate the meta-paths for comparative experiments. This algorithm better preserves the contextual information, including the correlation and differentiation between nodes, compared to traditional heterogeneous network link prediction methods \cite{Fu2017HIN2Vec}. Jiang et al. improved the graph convolution neural network, which firstly replaces the traditional graph convolution's transfer layer by a HeGCN layer to make it more suitable for dealing with heterogeneous networks and combines with an adversarial learning layer to obtain node characterization, and then the Adam model is is trained to reduce the value of the loss function, which improves the AUC metrics and PA metrics in link prediction compared to the deep walk and Hin2vec algorithms \cite{Jiang2022LinkPrediction}.

Edges in directed networks carry directional information, and the directionality of edges is very common in real-world problems, and ignoring the directional information in them will reduce the model predictability. For link prediction in directed networks, Lichtenwalter et al. proposed the PropFlow algorithm based on random walk algorithm, which assigns parameters to potential connections through a baseline, and the objects will choose routes based on the weights when walking while restricting the direction of the object's walking to the direction of the edges, and the objects do not need to restart or convergence, so the computation time can be reduced \cite{Lichtenwalter2010LinkPrediction}. Ramesh Sarukkai et al. used Markov chain algorithm in machine learning for link prediction and path research, which can be modeled dynamically according to the historical change condition and combined with eigenvector decomposition to obtain link tours, and this algorithm provides an effective tool for the research of Web path analysis and link prediction \cite{Sarukkai2000MarkovChains}. Pan et al. applied neural network to the link prediction of directed graph proposed NNLP algorithm, more suitable and combined with the standard particle swarm algorithm to optimize it, and use error back propagation algorithm and BP iterative learning algorithm to train neural network, the training process fusion using CN, AA, PA, Katz and other indicators, thus improving the model's complementarity and accuracy \cite{Pan2018LinkPredictionNN}.

In addressing the evolving landscape of link prediction, our study employs the same rigorous evaluation metrics as previous research, to ensure comparability and consistency. We have established a robust baseline by encompassing heuristic-based methods and machine learning approaches, which sets the stage for our advanced GNN models. 

% \begin{quote}
% This is an example of a quote.
% \end{quote}

%%%%%%%%%%%%%%%%%%%%%%%%%%%%%%%%%%%%%%%%%%
\section{Datasets}
\subsection{Data Collection}
In this study, we harnessed the Web of Science database to construct a dataset central to the technological advances in zinc batteries, an area increasingly vital due to the surge in demand for green energy solutions and new energy electric vehicles. Recognizing zinc's relative abundance compared to lithium, and its potential for industrial application, our search targeted a corpus reflective of the most recent and impactful research. 

We retrieved a comprehensive collection of 10,187 research articles using a refined search strategy based on relevant keywords that signify the core technologies and innovations in the field: TI=(Zinc) AND TI=(Anode) OR TI=(Zn) AND TI=(Anode) OR TI=(Zn) AND TI=(Batter*) OR TI=(Zinc) AND TI=(Batter*) \cite{shi2020overview}. This strategy was meticulously designed to capture the breadth and depth of contemporary zinc battery research, which is predominantly represented in publications post-2010, accounting for over 90\% of the total corpus and indicating a burgeoning interest and development in this domain. The dataset includes articles until November 2023, with a significant concentration of papers from authoritative journals such as the Journal of Power Sources, Chemical Engineering Journal, ACS Applied Materials \& Interfaces, Journal of the Electrochemical Society, and Journal of Materials Chemistry A. This approach ensures that our analysis is grounded in the most current and influential studies, providing a robust foundation for our subsequent predictive modeling and network analysis.

\subsection{Train-Test Split}
In our study, a network graph was generated from an edge dataset, representing the connections between nodes. This graph was then divided into training and test sets, adhering to a standard 80:20 split ratio, resulting in a training set comprising 80\% of the data and a test set with the remaining 20\%. 

To counter the issue of class imbalance typically present in network data, we employed a novel approach to generate synthetic non-links. This procedure was continued until the number of non-links equaled the number of actual links within the training set, ensuring a balanced distribution of classes. This method was mirrored in the preparation of the test set. As a result, the training set expanded to include 296,574 samples, while the test set contained 74,144 samples. This deliberate oversampling strategy for the training set was critical for developing a model that can accurately predict both the existence and absence of links, which is essential for robust network analysis.

\section{Methods}

% \section{Experiments}
\subsection{Heuristic-Based Methods}

\subsubsection{Overview}

% \begin{table}[H] 
% \caption{This is a table caption. Tables should be placed in the main text near to the first time they are~cited.\label{tab1}}
% \newcolumntype{C}{>{\centering\arraybackslash}X}
% \begin{tabularx}{\textwidth}{CCC}
% \toprule
% \textbf{Title 1}	& \textbf{Title 2}	& \textbf{Title 3}\\
% \midrule
% Entry 1		& Data			& Data\\
% Entry 2		& Data			& Data\\
% \bottomrule
% \end{tabularx}
% \end{table}
% \unskip

% \begin{table}[htbp]
% \centering
% \caption{Heuristic-Based Methods Overview}
% \label{tab:heuristic_methods}
% \begin{tabular}{lccc}
% \toprule
% \textbf{Method} & \textbf{Accuracy} & \textbf{Precision} & \textbf{Recall} & \textbf{F1 Score} & \textbf{AUC Score} \\
% \midrule
% Random Prediction & 0.5 & 0.5 & 0.5 & 0.5 & 0.5 \\
% Common Neighbor & 0.854 & 0.844 & 0.869 & 0.856 & 0.854 \\
% Jaccard Coefficient & 0.826 & 0.854 & 0.786 & 0.819 & 0.826 \\
% Adamic/Adar & 0.806 & 0.942 & 0.651 & 0.77 & 0.806 \\
% Resource Allocation & 0.847 & 0.914 & 0.766 & 0.834 & 0.847 \\
% \bottomrule
% \end{tabular}
% \end{table}

Table \ref{tab:heuristic_methods} provides a summarized comparison of various heuristic-based methods for link prediction in terms of their performance metrics: Precision, Recall, F1 Score, and AUC Score. The table lists five methods, from a baseline 'Random Prediction' to more sophisticated approaches like 'Resource Allocation'. Each method's efficacy is quantified, with 'Adamic/Adar' showing the highest precision and 'Common Neighbor' exhibiting the best balance between precision and recall as reflected by its F1 Score.

\begin{table}[htbp]
\centering
\caption{Heuristic-Based Methods Overview}
\label{tab:heuristic_methods}
\newcolumntype{C}{>{\centering\arraybackslash}X}
\begin{tabularx}{\textwidth}{CCCCC}
\toprule
\textbf{Method} & \textbf{Precision} & \textbf{Recall} & \textbf{F1 Score} & \textbf{AUC Score} \\
\midrule
Random Prediction & 0.5 & 0.5 & 0.5 & 0.5 \\
Common Neighbor & 0.844 & 0.869 & 0.856 & 0.854 \\
Jaccard Coefficient & 0.854 & 0.786 & 0.819 & 0.826 \\
Adamic/Adar & 0.942 & 0.651 & 0.77 & 0.806 \\
Resource Allocation & 0.914 & 0.766 & 0.834 & 0.847 \\
\bottomrule
\end{tabularx}
\end{table}

% \begin{listing}[H]
% \caption{Title of the listing}
% \rule{\columnwidth}{1pt}
% \raggedright Text of the listing. In font size footnotesize, small, or normalsize. Preferred format: left aligned and single spaced. Preferred border format: top border line and bottom border line.
% \rule{\columnwidth}{1pt}
% \end{listing}

\subsubsection{Random Prediction}
A random prediction method is employed as a baseline model for evaluating the test set. This method involves randomly assigning binary labels (0 or 1) to each element of the test set, effectively simulating a naive approach where predictions are made without any data-driven learning or inference. This technique establishes a foundational benchmark for assessing the efficacy of more sophisticated models.

The outcomes of this random prediction strategy are quantified using standard performance metrics. The results demonstrate equal values across all metrics with a precision, recall, F1 score, and AUC (Area Under the Curve) score, all at approximately 0.5002. These metrics indicate that the model performs at a level akin to random guessing, as expected from this method. The consistency across all metrics underscores the non-discriminatory nature of the random prediction approach. These results set a baseline that more advanced, data-informed models must exceed to be considered valuable in predictive analytics within this domain.

\subsubsection{Common Neighbors}
In this segment of the analysis, the Common Neighbors Method is implemented as a technique for link prediction. This method operates on the principle that two nodes in a network are more likely to form a link if they share a significant number of common neighbors. The approach involves calculating the number of common neighbors for each pair of nodes in the test set, and a prediction of a link (label 1) is made if the common neighbors count is one or more; otherwise, no link (label 0) is predicted.

The performance of the Common Neighbors Method is evaluated through several metrics, demonstrating notable effectiveness in predicting links. The precision, which measures the correctness of predicted links, is 0.8437, suggesting that a significant proportion of links predicted by the model are actual links. The recall of 0.8691 implies that the method is effective in identifying actual links within the test set. The F1 score, a balance between precision and recall, stands at 0.8562, further attesting to the method's robustness. Finally, the AUC score of 0.8540 indicates a strong ability to differentiate between the presence and absence of links. These results suggest that the Common Neighbors Method is a reliable tool in network analysis for link prediction, substantially surpassing the baseline established by the random prediction approach.

\subsubsection{Jaccard Coefficient}
The Jaccard Coefficient Method \cite{Jaccard1912}, another approach for link prediction, is based on the similarity between the sets of neighbors of two nodes. The Jaccard coefficient is calculated as the ratio of the size of the intersection to the size of the union of the neighbor sets of two nodes. Predictions are made based on a set threshold; in this case, a link is predicted if the Jaccard score is equal to or exceeds 0.01.

The performance metrics for the Jaccard Coefficient Method show its effectiveness in link prediction. The precision is 0.8537, suggesting that a significant portion of the predicted links are true links. The recall of 0.7862 demonstrates the method's capability in identifying a majority of actual links. The F1 score, at 0.8186, reflects a balanced measure of precision and recall. Additionally, an AUC score of 0.8257 highlights the method's proficiency in distinguishing between link presence and absence. These results showcase the Jaccard Coefficient Method as a viable and efficient tool for predicting links in network analysis.

\subsubsection{Adamic/Adar Index}

The Adamic/Adar Index Method \cite{Adamic2003} is implemented for link prediction, focusing on the shared neighbors of two nodes. The method computes the Adamic/Adar index, which accounts for the number of common neighbors and inversely weights these neighbors by the logarithm of their degrees. A threshold is set (0.5 in this case) to decide whether a link is predicted based on the index value.

The method's performance is summarized as follows: it achieves a precision of 0.9421, a recall of 0.6514, an F1 score of 0.7702, and an AUC score of 0.8057. These metrics indicate that while the method is highly precise in its predictions, its recall is comparatively lower, suggesting a more conservative prediction of links. The overall results demonstrate the method's efficiency in discerning link presence in the network.

\subsubsection{Resource Allocation Index}

The Resource Allocation Index Method \cite{Zhou2009}, applied here for link prediction, is based on the idea of allocating resources through common neighbors between pairs of nodes. The Resource Allocation (RA) index is calculated for each pair in the test set, taking into account the degree of common neighbors. A predefined threshold (0.01 in this case) is used to determine if a link is predicted based on the RA index score.

The performance metrics for the Resource Allocation Index Method are as follows: a precision of 0.9144, indicating a high degree of accuracy in the predicted links; a recall of 0.7662, showing the method's ability to identify a substantial portion of actual links; and an F1 score of 0.8337, reflecting a balanced trade-off between precision and recall. The AUC score is 0.8472, suggesting a strong discriminatory ability between link presence and absence.

\subsection{Machine Learning Methods}

\subsubsection{Overview}
Table \ref{tab:ml_methods} delineates the performance of three machine learning algorithms—Logistic Regression, Random Forest, and XGBoost—on link prediction tasks, evaluated by Precision, Recall, F1 Score, and AUC Score. It highlights that while XGBoost achieves the highest precision, Random Forest outperforms the others in terms of the F1 Score, suggesting a more balanced performance in precision and recall.

\begin{table}[htbp]
\centering
\caption{Machine Learning Methods Overview}
\label{tab:ml_methods}
\newcolumntype{C}{>{\centering\arraybackslash}X}
\begin{tabularx}{\textwidth}{CCCCC}
\toprule
\textbf{Method} & \textbf{Precision} & \textbf{Recall} & \textbf{F1 Score} & \textbf{AUC Score} \\
\midrule
Logistic Regression & 0.78 & 0.709 & 0.743 & 0.754 \\
Random Forest & 0.838 & 0.793 & 0.815 & 0.82 \\
XGBoost & 0.864 & 0.635 & 0.794 & 0.809 \\
\bottomrule
\end{tabularx}
\end{table}

\subsubsection{Feature Engineering}

The feature engineering phase in our study is a pivotal step towards enabling effective network analysis. Initially, we standardize numerical data using a standard scaler to ensure mean normalization and variance scaling. For categorical variables, a one-hot encoding scheme \cite{harris2012digital} is applied to convert these variables into a binary vector representation, avoiding any implicit ordering that could be mistakenly inferred by the models. Textual data undergoes a transformation via TF-IDF (Term Frequency-Inverse Document Frequency) vectorization \cite{sparckjones1972statistical}, a technique that reflects the importance of words relative to a document collection, hence allowing for the nuanced representation of text features.

The culmination of these preprocessing techniques is a multi-dimensional feature space, structured as a sparse matrix to manage the high dimensionality efficiently. The final feature space encompasses a total of 10,187 samples, each characterized by a feature vector consisting of 2,026 distinct elements. This feature space's dimensionality was carefully curated to capture the rich, underlying structures of the node attributes while avoiding the pitfalls of overfitting that can occur with excessively high-dimensional data. The resulting dataset, formed through careful curation and transformation of node characteristics, provides a comprehensive and robust foundation for the subsequent phases of predictive modeling in our network analysis.
% \begin{table}[htbp]
% \centering
% \caption{Heuristic-Based Methods Overview}
% \label{tab:heuristic_methods}
% \newcolumntype{Y}{>{\centering\arraybackslash}X} % centering with tabularx
% \begin{tabularx}{\textwidth}{lYYYYY} % 'l' for the method column, 'Y' for the rest
% \toprule
% \textbf{Method} & \textbf{Accuracy} & \textbf{Precision} & \textbf{Recall} & \textbf{F1 Score} & \textbf{AUC Score} \\
% \midrule
% Logistic Regression & 0.754 & 0.78 & 0.709 & 0.743 & 0.754 \\
% Random Forest & 0.82 & 0.838 & 0.793 & 0.815 & 0.82 \\
% XGBoost & 0.809 & 0.864 & 0.635 & 0.794 & 0.809 \\
% \bottomrule
% \end{tabularx}
% \end{table}
% \unskip

\subsubsection{Logistic Regression}

In this study, a Logistic Regression model \cite{Berkson1944} is employed as a machine learning method for link prediction in the network. Logistic Regression, a widely used statistical model for binary classification, is trained on the engineered feature set, which encapsulates the characteristics of node pairs in the network. The model is optimized to predict the likelihood of link existence between nodes based on the features derived from their attributes.

The performance of the Logistic Regression model is evaluated using standard metrics on both training and test datasets. In the training phase, the model achieves a precision of 0.7242, a recall of 0.7220, an F1 score of 0.7231, and an AUC score of 0.7235. For the test set, the model demonstrates a precision of 0.7799, a recall of 0.7090, an F1 score of 0.7428, and an AUC score of 0.7544. These metrics indicate the model's effectiveness in accurately predicting links, with a notable performance improvement in the test set compared to the training set. This suggests that the Logistic Regression model, trained on the comprehensive feature set, is a robust tool for link prediction in network analysis.

\subsubsection{Random Forest}

In this phase of the study, a RandomForestClassifier, a powerful ensemble learning method, is utilized for link prediction in the network. The RandomForest model \cite{Ho1995} \cite{Ho1998}, known for its robustness and ability to handle complex datasets, is trained on the prepared feature set, which includes a diverse range of node attributes and their interactions.

The RandomForest model's performance is assessed using a variety of metrics. During training, the model demonstrates exceptionally high precision, recall, F1 score, and AUC score, all approximately 0.9986, indicating an almost perfect fit to the training data. However, when evaluated on the test set, the model exhibits a precision of 0.8379, a recall of 0.7926, an F1 score of 0.8146, and an AUC score of 0.8196. These results reflect the model's effectiveness in generalizing to new data, with a notable distinction between its training and testing performance. The high precision on the test set suggests the model's strong capability in accurately identifying true links, while the recall indicates its efficiency in capturing a substantial proportion of actual links. These metrics collectively demonstrate the RandomForest model's efficacy as a robust tool for predicting links in network analysis.

\subsubsection{XGBoost}
The study employs the XGBoost classifier, an advanced gradient boosting framework, for the task of link prediction in networks. XGBoost \cite{ChenGuestrin2016} is renowned for its high efficiency and effectiveness, particularly in dealing with complex datasets. It operates by sequentially building trees, where each new tree attempts to correct the errors made by the previous ones, resulting in a strong predictive model.

In terms of performance, the XGBoost model demonstrates a balanced and effective capability in both the training and testing phases. The training metrics reveal an accuracy of 0.7948, precision of 0.8186, recall of 0.7574, F1 score of 0.7869, and an AUC score of 0.7948. The test metrics exhibit an accuracy of 0.8095, precision of 0.8637, recall of 0.7349, F1 score of 0.7941, and an AUC score of 0.8095. These results indicate the model's strong performance in accurately predicting links, with particularly high precision in the test set, suggesting its ability to generalize well to new data. The XGBoost model thus proves to be a robust and reliable tool in the context of network analysis for link prediction.

\subsection{Graph Neural Networks}
\begin{figure}[H]

\centering
\includegraphics[width=10.5cm]{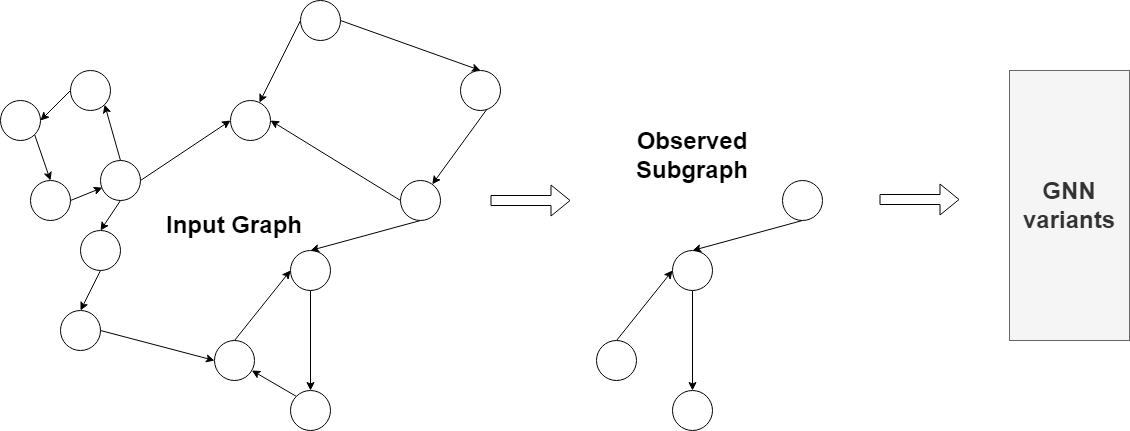}
\caption{The general methodology of different GNN variants. \label{gnn_process}}
\end{figure}

Figure \ref{gnn_process} illustrates the process by which the GNN models utilize subgraphs observed from the input graph to predict links. This visual representation clarifies the methodology underlying the application of various GNN variants and demonstrates the nuanced approach of our analysis.

\subsubsection{Overview}
Table \ref{tab:gnn_methods} shows the comparison between Graph Neural Network (GNN) models with and without the integration of Louvain community detection reveals the impact of incorporating structural community information into link prediction tasks. The results indicate a consistent improvement in model performance when community detection is applied.

For the GAT model, the addition of Louvain community detection leads to increases in precision, recall, F1 score, and AUC score, suggesting that community context enhances the model's ability to predict links accurately. Similar trends are observed in the GATv2 model, where integrating community features results in a rise in all evaluated metrics, notably in the AUC score.

The GCN model, already exhibiting strong performance, further benefits from the inclusion of community detection, as evidenced by the increase in precision, recall, F1 score, and AUC score. This improvement is also mirrored in the GCNv2 model, although the precision slightly decreases while the other metrics show enhancements.

Lastly, the GraphSAGE model shows considerable improvements across all metrics with the inclusion of Louvain community detection. This suggests that adding community features is particularly effective for models like GraphSAGE, which rely heavily on neighborhood aggregation.

\begin{table}[htbp]
\centering
\caption{Graph Neural Networks Overview}
\label{tab:gnn_methods}
\newcolumntype{C}{>{\centering\arraybackslash}X}
\begin{tabularx}{\textwidth}{CCCCC}
\toprule
\textbf{Method} & \textbf{Precision} & \textbf{Recall} & \textbf{F1 Score} & \textbf{AUC Score} \\
\midrule
GAT             & 0.573 & 0.918 & 0.705 & 0.777 \\
GAT + Louvain   & 0.606 & 0.923 & 0.732 & \textbf{0.823} \\
GATv2           & 0.582 & 0.932 & 0.717 & 0.812 \\
GATv2 + Louvain & 0.608 & 0.913 & 0.730 & \textbf{0.830} \\
GCN             & 0.664 & 0.957 & 0.784 & 0.892 \\
GCN + Louvain   & 0.682 & 0.965 & 0.799 & \textbf{0.906} \\
GCNv2           & 0.700 & 0.976 & 0.815 & 0.917 \\
GCNv2 + Louvain & 0.680 & 0.970 & 0.799 & \textbf{0.919} \\
GraphSAGE       & 0.656 & 0.939 & 0.772 & 0.859 \\
GraphSAGE + Louvain & 0.688 & 0.952 & 0.799 & \textbf{0.878} \\
\bottomrule
\end{tabularx}
\end{table}

The confusion matrices presented in figure \ref{confusion_matrices} correspond to the models listed in table, with the same ordering. They illustrate the performance of each model on a consistent test set, with a fixed random seed of 42 to ensure reproducibility. The left side of each matrix pertains to models using only the original node features, while the right side shows the performance after the integration of community detection features via the Louvain method.

\subsubsection{Graph Attention Networks}

In this study, a Graph Attention Network (GAT) is employed for link prediction, leveraging the power of attention mechanisms in graph neural networks. The GAT model uses multiple attention heads to focus on different parts of the node's neighborhood, allowing for a more nuanced aggregation of neighbor information. This structure is particularly effective in capturing the complex dependencies within the graph data.

The GAT model achieves a precision of 0.6065 and a recall of 0.9231 in link prediction, with an F1 score of 0.7320 and an AUC score of 0.8225. These metrics underscore its effectiveness in identifying relevant links and its strong overall predictive performance.

The study employs GATv2 \cite{Brody2022}, an evolution of the Graph Attention Network (GAT), which incorporates enhanced attention mechanisms for link prediction. GATv2 is characterized by its attention heads that enable more nuanced feature aggregation from a node's neighborhood, effectively capturing complex interactions within the graph.

In testing, the GATv2 model demonstrates a precision of 0.5824 and a recall of 0.9317, culminating in an F1 score of 0.7168. The AUC score stands at 0.8118, indicating its proficient predictive ability in distinguishing between the presence and absence of links.

\begin{figure}[H]

\centering
\includegraphics[width=10.5cm]{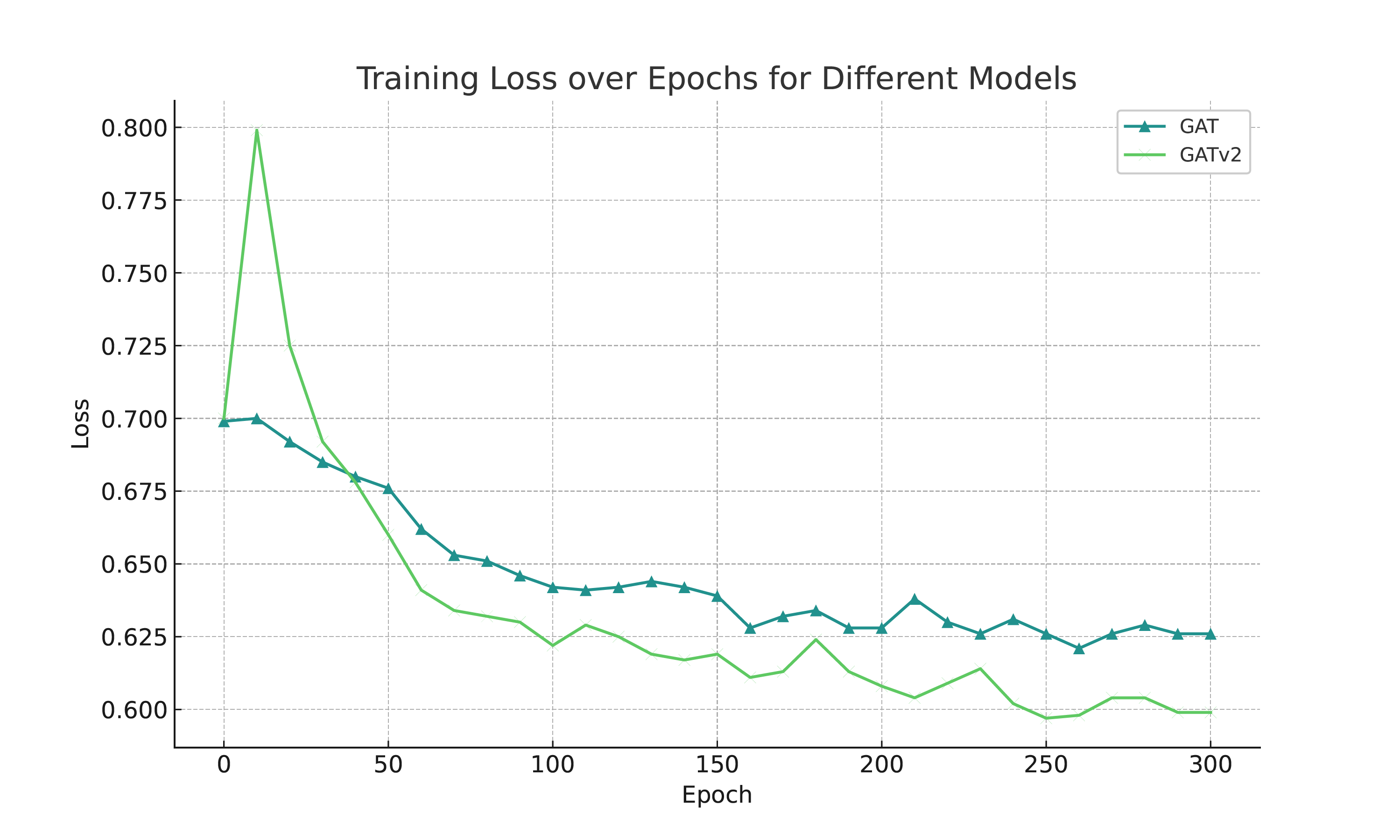}
\caption{Training loss over epochs for GAT models. \label{gat}}
\end{figure}

Figure \ref{gat} presents the training loss trajectories for both the GAT and GATv2 models over a span of 300 epochs, providing a visual representation of the models' learning progress. The GAT model's loss, denoted by triangles, shows an initial rapid decline, stabilizing as the epochs increase, which is indicative of the model effectively learning from the training data. Similarly, the GATv2 model, begins with a higher loss that quickly descends, albeit with some early fluctuations, before it too stabilizes. This pattern is reflective of the GATv2's advanced attention mechanisms adapting to the graph structure over time. The colors selected for the plot are consistent with those used in previous figures for ease of comparison.

In our experimental validation, the hyperparameters for the GAT and GATv2 model were optimized through comprehensive testing. The final configuration, yielding the most efficacious results, is delineated in table \ref{GAT parameter} and \ref{GATv2 parameter}.

\subsubsection{Graph Convolutional Networks}
In this segment of the research, a Graph Convolutional Network (GCN) is implemented for link prediction in network analysis. This GCN includes additional layers and dropout for regularization, enhancing its capability to capture complex patterns in graph data. The model, by operating directly on the graph structure, efficiently aggregates and transforms node features, leveraging the inherent topological information.

The GCN model demonstrates commendable performance in link prediction, achieving a precision of 0.6638, a recall of 0.9570, and an F1 score of 0.7839. Its AUC score of 0.8921 highlights its effective discrimination between link presence and absence, underscoring the model's robustness in network analysis.

The GCNv2 model \cite{Tang2019}, an advanced version of the Graph Convolutional Network (GCN), is implemented for link prediction. The GCNv2 model integrates additional convolution layers and a linear transformation layer, enhancing its capacity to process complex graph structures. Key parameters such as alpha and theta are adjusted to optimize the model's performance.

The GCNv2 model exhibits impressive results in link prediction, achieving a precision of 0.6796, a recall of 0.9697, and an F1 score of 0.7991. Its AUC score of 0.9199 further demonstrates its strong capability in differentiating between the presence and absence of links in the network.

\begin{figure}[H]

\centering
\includegraphics[width=10.5cm]{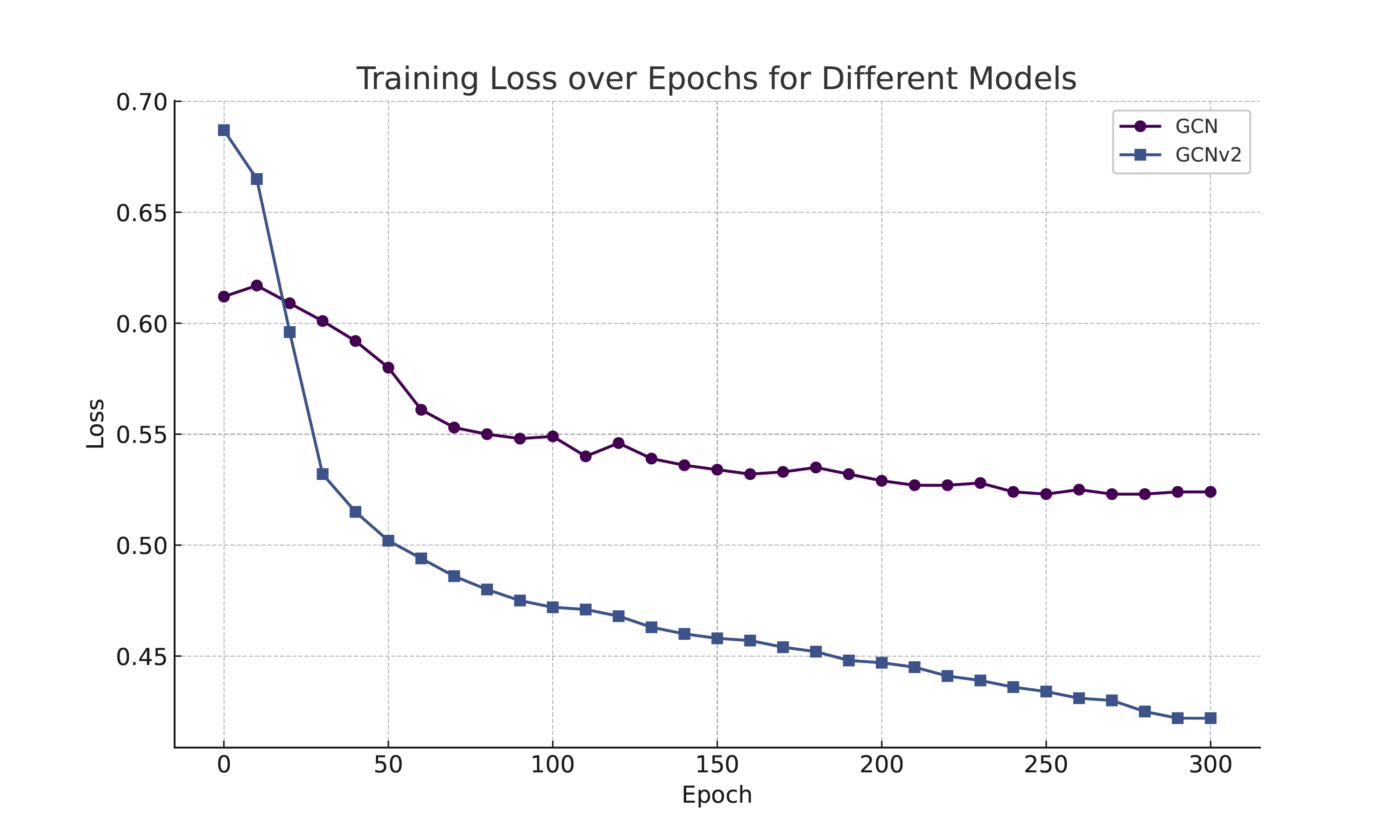}
\caption{Training loss over epochs for GCN models. \label{gcn}}
\end{figure}

Figure \ref{gcn} displays the training loss trajectories for the GCN and GCNv2 models across 300 training epochs. The GCN model's losses are indicated with circles and exhibit a gradual descent with some variability, suggesting an incremental learning process. In contrast, the GCNv2 model, marked with squares, shows a steeper initial decline in loss, indicating a rapid adaptation to the data. This steep decline in the early epochs for GCNv2 can be attributed to the enhanced convolution layers and linear transformation layer that allow it to capture complex graph structures more effectively.

The hyperparameter tuning undertaken for the GCN and GCNv2 model was systematic, aiming to ascertain an optimal set of configurations for enhancing model performance. The resulting parameters, determined to be most effective through empirical testing, are in table \ref{GCN parameter} and \ref{GCNv2 parameter}.

\subsubsection{GraphSAGE}
The study implements GraphSAGE, a Graph Neural Network model known for its effectiveness in learning from large graphs. GraphSAGE utilizes neighborhood sampling and aggregating functions to generate node embeddings, efficiently capturing local graph structures.

The GraphSAGE model demonstrates a precision of 0.6556 and a recall of 0.9392 in link prediction, resulting in an F1 score of 0.7722. Its AUC score of 0.8589 indicates a strong capability in distinguishing between the presence and absence of links.

\begin{figure}[H]

\centering
\includegraphics[width=10.5cm]{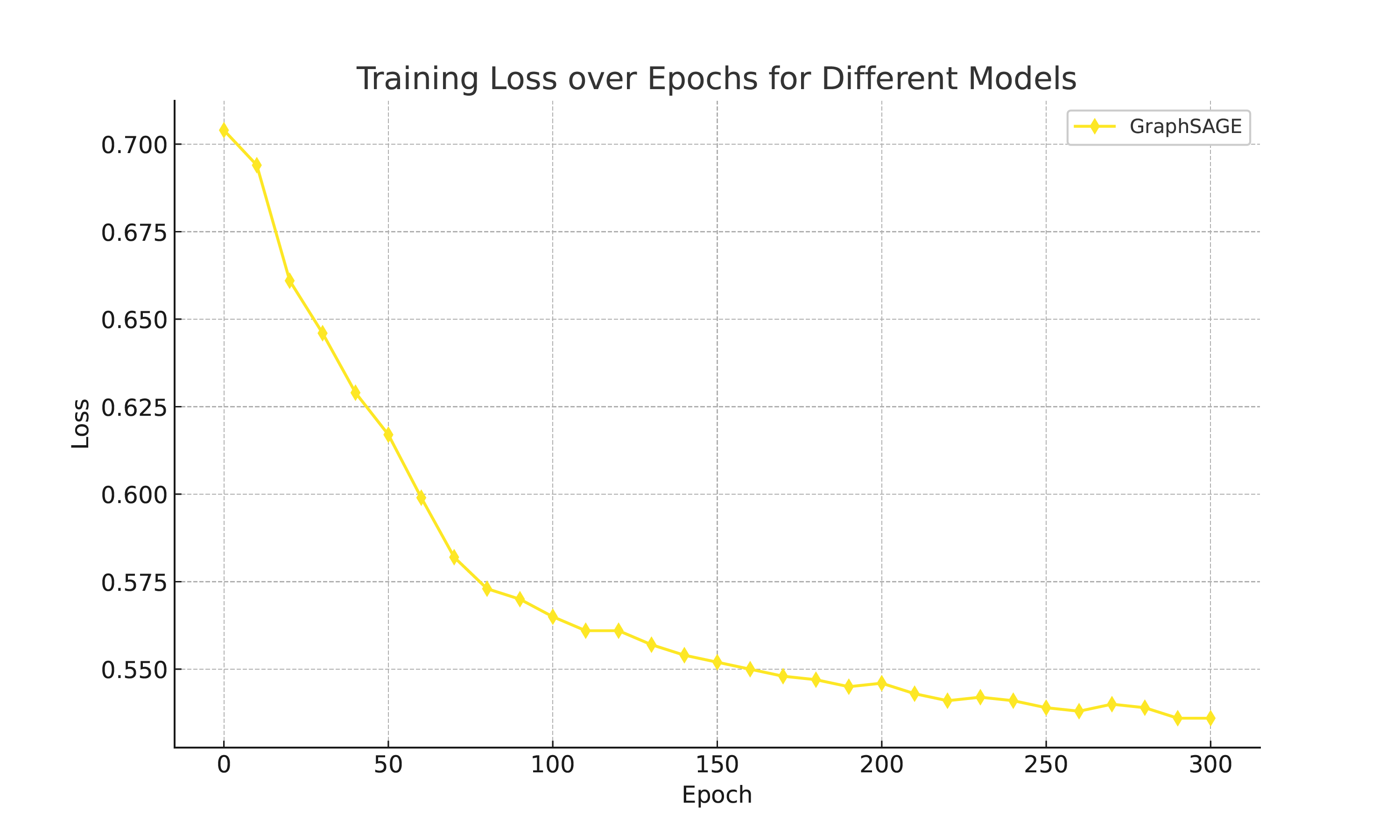}
\caption{Training loss over epochs for GraphSAGE model. \label{graphsage}}
\end{figure}

Figure \ref{graphsage} presents the training loss trajectory for the GraphSAGE model over 300 epochs, providing insight into the model's learning performance throughout the training phase. The loss curve, depicted with a yellow line, begins with a sharp decline, indicating a rapid initial learning phase, and then transitions into a more gradual descent. This suggests that GraphSAGE quickly assimilates the local graph structures and then refines its understanding of the data over time. The relative stability of the loss after the initial epochs showcases the efficiency of GraphSAGE's neighborhood sampling and aggregation functions in generating informative node embeddings.

The hyperparameter tuning process was executed with the goal of identifying the most effective model parameters. The selected hyperparameters for the Enhanced GraphSAGE model are detailed in the table \ref{GraphSAGE parameter}.

\subsubsection{Graph Neural Networks with Community Detection}

In the enhanced methodology of our study, the preprocessing of graph data is elevated through the integration of the Louvain community detection algorithm \cite{blondel2008fast}. The Louvain method operates on the principle of modularity optimization. Mathematically, it seeks to maximize the modularity \( Q \), defined as 

\[ Q = \frac{1}{2m} \sum_{ij} \left[ A_{ij} - \frac{k_i k_j}{2m} \right] \delta(c_i, c_j), \]

where \( A \) is the adjacency matrix of the network, \( m \) is the sum of all edge weights, \( k_i \) and \( k_j \) are the degrees of nodes \( i \) and \( j \), and \( \delta \) is the Kronecker delta function that equals 1 if nodes \( i \) and \( j \) are in the same community and 0 otherwise.

By applying this algorithm, each node is endowed with a community label that encapsulates its modular connectivity within the graph. These labels are then encoded into a one-hot vector, expanding the original node feature set—previously vectorized by TF-IDF and one-hot encoding techniques—by appending this new community structure information. Consequently, the feature space dimensions increase from 2,026 to 3,315, enriching the input data for the GNN models.

\begin{figure}[H]

\centering
\includegraphics[width=10.5cm]{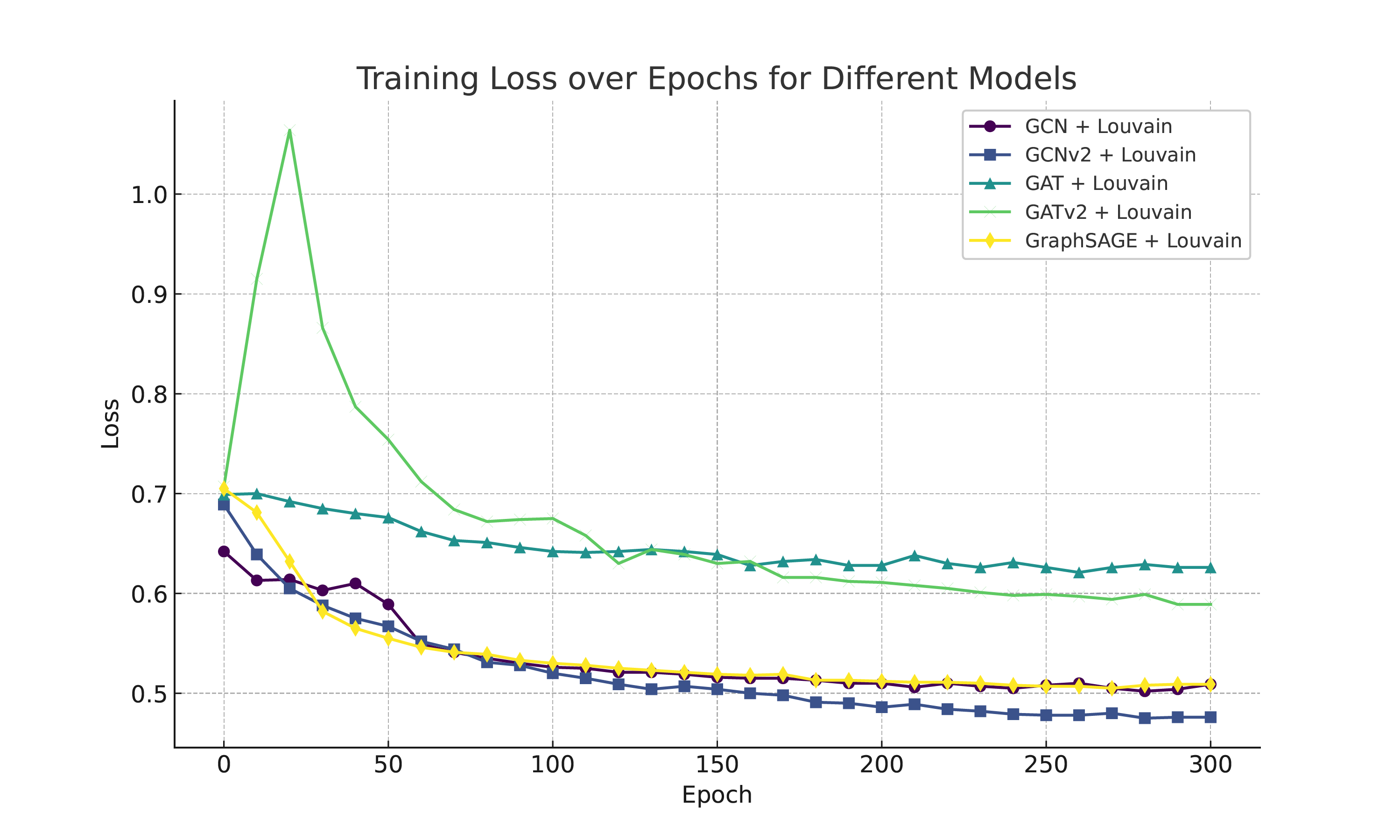}
\caption{Training loss over epochs for different models. \label{fig1}}
\end{figure}

The GNN models—GCN, GCNv2, GAT, GATv2, and GraphSAGE—are subsequently trained on this augmented dataset, harnessing not only the intrinsic node features but also the macro-structural properties encoded by the Louvain method. This dual representation fosters a deeper interpretative capacity within the models, leading to an improved learning trajectory as observed in Figure ~\ref{fig1}. The loss trajectories for models trained with community detection features indicate a more pronounced and rapid stabilization, highlighting the effectiveness of the Louvain method in enhancing the network analysis capabilities of GNNs.

\section{Discussion}

%%%%%%%%%%%%%%%%%%%%%%%%%%%%%%%%%%%%%%%%%%
\subsection{Limitation}
The Louvain community detection algorithm, when applied in Graph Neural Networks for link prediction, may encounter a resolution limit issue. This limitation arises because the algorithm is prone to merging smaller communities into larger ones, thereby potentially overlooking finer, yet significant, community structures. This oversight can lead to suboptimal performance in link prediction tasks where the identification of smaller, more nuanced communities is crucial for accurate predictions.

\subsection{Future Work}

As we look to the future, the refinement of integrating community detection with GNNs stands as a promising avenue for advancing link prediction techniques. Beyond simply enhancing the accuracy of link predictions, the amalgamation of these two methodologies has the potential to significantly influence the broader field of network analysis. The nuanced understanding of community structures within networks could revolutionize the way we approach complex networked systems, enabling a deeper comprehension of their evolution, resilience, and underlying dynamics.

In the realm of scientific literature, the implications of this research are manifold. By improving paper recommendation systems, our approach can streamline the research process, enabling scholars to discover relevant works with unprecedented precision. This is particularly vital in an era where the sheer volume of publications can overwhelm researchers. Moreover, by anticipating collaboration and citation patterns, we can not only track but also predict the progression of scientific discourse, potentially catalyzing new research intersections and collaborations that might have otherwise remained undiscovered.

The integration of community detection with GNNs could also be extended to other domains where network analysis plays a critical role, such as social network analysis, epidemiology, and bioinformatics. For instance, in social networks, this methodology could enhance the detection of community structures and their changes over time, providing insights into the formation of social groups and the spread of information or misinformation. In epidemiology, it can aid in understanding and predicting the spread of diseases within and between communities, potentially informing public health strategies and interventions. In bioinformatics, it could improve the prediction of protein-protein interactions, which is crucial for understanding cellular processes and the development of drugs.

Furthermore, exploring alternative community detection algorithms can address current limitations and open new research directions, such as the study of community evolution in dynamic networks. The potential to adapt GNN models to better incorporate temporal and structural changes within communities could lead to groundbreaking developments in dynamic network analysis.

In summary, the integration of community detection methods with GNN architectures represents not just a methodological enhancement in link prediction tasks but also a strategic shift towards a more sophisticated understanding of networks. This could lead to developing more intelligent systems that can adapt to and predict changes within networks, offering significant benefits across various scientific and social domains. Hence, our future work aims not only to improve upon the technical aspects of our approach but also to explore its broader applications, thereby contributing to the progressive evolution of network science and its capacity to address complex real-world problems.

\section{Conclusion}
In conclusion, our study provides compelling quantitative evidence that the integration of the Louvain community detection method with various GNN models results in robust performance enhancements in link prediction tasks. For instance, when Louvain is applied to GNN models, we observe significant improvements in AUC scores, such as an increase from 0.777 to 0.823 for GAT and from 0.892 to 0.917 for GCNv2. These improvements are consistent across different GNN architectures, underlining the efficacy of incorporating community structures into predictive modeling. 

Furthermore, our approach of Louvain-enhanced GNNs not only consistently outperforms traditional machine learning methods, as evidenced by the highest AUC score of 0.917 achieved by GCNv2 with Louvain, but also demonstrates superior results compared to heuristic-based methods, with the Resource Allocation index achieving an AUC score of 0.847. Such findings advocate for a paradigm shift towards the adoption of community-aware GNN models for link prediction, highlighting the importance of a synergistic perspective that blends node-specific features with macroscopic network structures. This study, therefore, concludes that leveraging the latent community information within GNN architectures can substantially elevate the performance of link prediction models, setting a new benchmark for future research in this domain.

% This section is not mandatory, but can be added to the manuscript if the discussion is unusually long or complex.

% %%%%%%%%%%%%%%%%%%%%%%%%%%%%%%%%%%%%%%%%%%
% \section{Patents}

% This section is not mandatory, but may be added if there are patents resulting from the work reported in this manuscript.

% %%%%%%%%%%%%%%%%%%%%%%%%%%%%%%%%%%%%%%%%%%
% \vspace{6pt} 

%%%%%%%%%%%%%%%%%%%%%%%%%%%%%%%%%%%%%%%%%%
%% optional
%\supplementary{The following supporting information can be downloaded at:  \linksupplementary{s1}, Figure S1: title; Table S1: title; Video S1: title.}


%%%%%%%%%%%%%%%%%%%%%%%%%%%%%%%%%%%%%%%%%%
\authorcontributions{Conceptualization, C.L. and Y.H.; methodology, Y.H and Y.S.; software, Y.H; validation, Y.H.; formal analysis, Y.H.; investigation, Y.H. and C.L.; data curation, C.L. and K.W.; writing---original draft preparation, Y.H. and C.L.; writing---review and editing, Y.H. and C.L.; visualization, Y.H.; supervision, C.L. and S.Y.; project administration, H.X.; All authors have read and agreed to the published version of the manuscript.}

\funding{This research was funded by the National Natural Science Funding of China grant number (No.72274113), in part by the Chinese Academy of Science’s “Light of West China” program and the Taishan Scholar Foundation of Shandong province of China (tsqn202103069).}

\institutionalreview{Not applicable.}

\informedconsent{Not applicable.}

\dataavailability{The data presented in this study are available on request from the corresponding author.}

\conflictsofinterest{The authors declare no conflict of interest.} 

%%%%%%%%%%%%%%%%%%%%%%%%%%%%%%%%%%%%%%%%%%
%% Optional

%% Only for journal Encyclopedia
%\entrylink{The Link to this entry published on the encyclopedia platform.}

\abbreviations{Abbreviations}{
The following abbreviations are used in this manuscript:\\

\noindent 
\begin{tabular}{@{}ll}
TF-IDF & Term frequency-inverse document frequency \\
XGBoost & Extreme gradient boosting \\
GNN & Graph neural network\\
GAT & Graph attention network\\
GCN & Graph convolutional network\\
GraphSAGE & Graph sample and aggregation\\
ReLU & Rectified linear unit\\
ELU & Exponential linear unit \\
\end{tabular}
}

%%%%%%%%%%%%%%%%%%%%%%%%%%%%%%%%%%%%%%%%%%
%% Optional
\appendixtitles{yes} % Leave argument "no" if all appendix headings stay EMPTY (then no dot is printed after "Appendix A"). If the appendix sections contain a heading then change the argument to "yes".
\appendixstart
\appendix
\section[\appendixname~\thesection]{Detailed Architecture and Hyperparameter Tuning of Enhanced GNN Models}

This appendix provides a detailed overview of the architecture of our Enhanced GNN models, as well as the hyperparameter tuning process that was undertaken to optimize their performance. These details are crucial for the reproducibility of our research and for providing insights into the potential of deep learning to enhance network analysis.

\subsection[\appendixname~\thesubsection]{Graph Attention Networks}
The architecture of our Enhanced GAT model is meticulously constructed to leverage the attention mechanism for node feature refinement. Beginning with the input layer, the model accepts node features and advances them through a graph attentional layer (conv1), which employs four attention heads to project the input into a higher dimensional space, here chosen as 64. Each head computes separate features which are then concatenated, allowing the model to attend to information from different representation subspaces. This layer is followed by an Exponential Linear Unit (ELU) activation function and a dropout layer with a rate of 0.6. The second graph attentional layer (conv2) consolidates the features from the previous layer into a 16-dimensional space. In this final layer, a single attention head is employed to focus on the most salient features for the link prediction task. The utilization of dropout before each attentional operation ensures that the model remains robust to overfitting.

\begin{table}[H]
\caption{Optimal Configuration for the Enhanced GAT Model\label{GAT parameter}}
\newcolumntype{C}{>{\centering\arraybackslash}X}
\begin{tabularx}{\textwidth}{CC}
\toprule
\textbf{Parameter} & \textbf{Value} \\
\hline
Learning Rate & 0.01 \\
Dropout Rate & 0.6 \\
Number of Attention Heads & 4 (conv1), 1 (conv2) \\
Number of Epochs & 300 \\
Optimizer & Adam \\
Loss Function & Binary Cross-Entropy (BCELoss) \\
\hline
\end{tabularx}
\end{table}

\subsection[\appendixname~\thesubsection]{Graph Attention Networks Version 2 (GATv2)}
The Enhanced GATv2 model is an iteration on the GAT architecture, refined to harness the second version of the graph attention mechanism. At its core, the input layer receives the node features and directs them through the first graph attentional layer (conv1), which utilizes four attention heads to expand the feature representation, subsequently concatenated to provide a rich, multidimensional feature space. Each attention head, operating with an individual weight matrix, allows the model to diversify the focus on different neighborhoods of the input graph, thereby enriching the feature extraction process. The conv1 layer outputs are then activated using an Exponential Linear Unit (ELU) to maintain a non-linear transformation while controlling for neuron saturation, followed by a dropout mechanism set at a rate of 0.6 to enhance model generalization.

Continuing the feature refinement, the architecture employs a second graph attentional layer (conv2) that consolidates the previously expanded features into a singular, 48-dimensional space conducive to the link prediction task. This layer uses a single attention head, concentrating the model's focus on the most pertinent features. As in the first layer, the dropout is applied post-activation to regularize the learning process.

The GATv2 model structure—starting with conv1 followed by ELU and dropout, and progressing through conv2 with the same sequence—presents an advanced approach to leveraging attention mechanisms for graph neural networks. This architecture is particularly adept at capturing and processing complex dependencies within graph data, a critical capability for predictive tasks in network analysis.

\begin{table}[H]
\caption{Optimal Configuration for the Enhanced GATv2 Model\label{GATv2 parameter}}
\newcolumntype{C}{>{\centering\arraybackslash}X}
\begin{tabularx}{\textwidth}{CC}
\toprule
\textbf{Parameter} & \textbf{Value} \\
\hline
Learning Rate & 0.02 \\
Dropout Rate & 0.6 \\
Number of Attention Heads & 4 (conv1), 1 (conv2) \\
% Hidden Layer Dimensionality & 48 \\
Number of Epochs & Determined by validation performance \\
Optimizer & Adam \\
Loss Function & Binary Cross-Entropy (BCELoss) \\
\hline
\end{tabularx}
\end{table}

\subsection[\appendixname~\thesubsection]{Graph Convolutional Networks}
The Enhanced GCN model's architecture is composed of a sequence of graph convolutional layers designed to systematically refine node features for optimal representation. The architecture initiates with an input layer that takes node features and subjects them to a graph convolutional operation (conv1), mapping them to a 512-dimensional hidden space. This is immediately succeeded by a ReLU activation function to introduce non-linearity, followed by a dropout operation with a rate of 0.5 to mitigate overfitting. The process proceeds to the second graph convolutional stage (conv2), which further processes the feature vector, reducing its dimensionality to 128. Similar to the previous stage, this is paired with a subsequent ReLU activation and dropout. The architecture is completed by a third graph convolutional layer (conv3), which provides an additional transformation, producing a 32-dimensional feature vector as output. The arrangement of the layers—conv1, ReLU, dropout, conv2, ReLU, dropout, and then conv3—facilitates a comprehensive feature learning mechanism, integral for the Enhanced GCN model's adeptness at link prediction tasks within network analysis.

\begin{table}[H]
\caption{Optimal Configuration for Our Enhanced GCN Model\label{GCN parameter}}
\newcolumntype{C}{>{\centering\arraybackslash}X}
\begin{tabularx}{\textwidth}{CC}
\toprule
\textbf{Parameter} & \textbf{Value} \\
\hline
Learning Rate & 0.01 \\
% Hidden Layer Dimensionality & 32 \\
Dropout Rate & 0.5 \\
Number of Epochs & 300 \\
Optimizer & Adam \\
Loss Function & Binary Cross-Entropy (BCELoss) \\
\hline
\end{tabularx}
\end{table}

\subsection[\appendixname~\thesubsection]{Graph Convolutional Networks Version 2 (GCNv2)}
The Enhanced GCNv2 model introduces additional sophistication to the foundational GCN architecture by incorporating linear transformation layers and advanced convolutional operations. The input layer commences the process by receiving node features, which are then subjected to a linear transformation layer to reduce their dimensionality to 48. Following this, a sequence of GCNv2 convolutional layers (GCN2Conv) are applied. These layers are equipped with parameters $alpha$ and $theta$, which are essential in controlling the mixture of the initial node features and the propagated features, as well as the trade-off between the two. Each GCN2Conv layer operates on the principle of iteratively refining node features while maintaining a connection to the original features, thus enabling a deeper level of feature extraction without over-smoothing.

The model's architecture is characterized by three successive GCN2Conv layers, each followed by a ReLU activation function to inject non-linearity, and a dropout layer with a rate of 0.5, serving as a regularizer. This structured progression of layers—linear transformation, followed by three iterations of GCN2Conv, ReLU, and dropout—provides a potent mechanism for feature learning that captures both local and global graph structures pertinent to link prediction.

\begin{table}[H]
\caption{Optimal Configuration for the Enhanced GCNv2 Model\label{GCNv2 parameter}}
\newcolumntype{C}{>{\centering\arraybackslash}X}
\begin{tabularx}{\textwidth}{CC}
\toprule
\textbf{Parameter} & \textbf{Value} \\
\hline
Learning Rate & 0.02 \\
Dropout Rate & 0.5 \\
Alpha & 0.3 \\
Theta & 0.7 \\
% Hidden Layer Dimensionality & 48 \
Number of Epochs & Defined by early stopping or maximum epochs \\
Optimizer & Adam \\
Loss Function & Binary Cross-Entropy (BCELoss) \\
\hline
\end{tabularx}
\end{table}

\subsection[\appendixname~\thesubsection]{Graph Sample and Aggregate Networks (GraphSAGE)}
The Enhanced GraphSAGE model's architecture is defined by a sequence of graph convolutional layers that are strategically structured to optimize the representation of node features for link prediction. The model begins with the first GraphSAGE convolutional layer (conv1), which takes the input node features and expands them to a 512-dimensional hidden layer. This is followed by a ReLU activation to introduce non-linearity, and a dropout layer with a rate of 0.5, designed to prevent overfitting. The model then proceeds to a second GraphSAGE convolutional layer (conv2) that reduces the dimensionality from 512 to 128, applying the same sequence of ReLU activation and dropout. The final layer (conv3) is an additional GraphSAGE convolutional layer, further condensing the features to a 32-dimensional hidden space. This progression—conv1, ReLU, dropout, followed by conv2, ReLU, dropout, and concluded with conv3—ensures a deep and comprehensive feature learning process, which is critical for capturing the complex patterns inherent in network data for effective link prediction.

\begin{table}[H]
\caption{Optimal Configuration for the Enhanced GraphSAGE Model\label{GraphSAGE parameter}}
\begin{tabularx}{\textwidth}{CC}
\toprule
\textbf{Parameter} & \textbf{Value} \\
\hline
Learning Rate & 0.01 \\
Dropout Rate & 0.5 \\
Hidden Layer Dimensionality & 32 \\
Number of Epochs & 300 \\
Optimizer & Adam \\
Loss Function & Binary Cross-Entropy (BCELoss) \\
\hline
\end{tabularx}
\end{table}

% \begin{table}[H] 
% \caption{This is a table caption.\label{tab5}}
% \newcolumntype{C}{>{\centering\arraybackslash}X}
% \begin{tabularx}{\textwidth}{CCC}
% \toprule
% \textbf{Title 1}	& \textbf{Title 2}	& \textbf{Title 3}\\
% \midrule
% Entry 1		& Data			& Data\\
% Entry 2		& Data			& Data\\
% \bottomrule
% \end{tabularx}
% \end{table}

\section[\appendixname~\thesection]{Analysis of Confusion Matrices for GNN Models}
The inclusion of confusion matrices in this study serves to provide a detailed assessment of the classification performance of various GNN models, both with and without the enhancement of the Louvain method. Confusion matrices offer a comprehensive visualization of the model's predictive capabilities, displaying the true positives, false positives, true negatives, and false negatives. This information is crucial for understanding the specific types of errors made by the models and for evaluating their performance beyond aggregated metrics like precision, recall, and F1 score.

\begin{figure}[H]

\centering
\includegraphics[width=13.5cm]{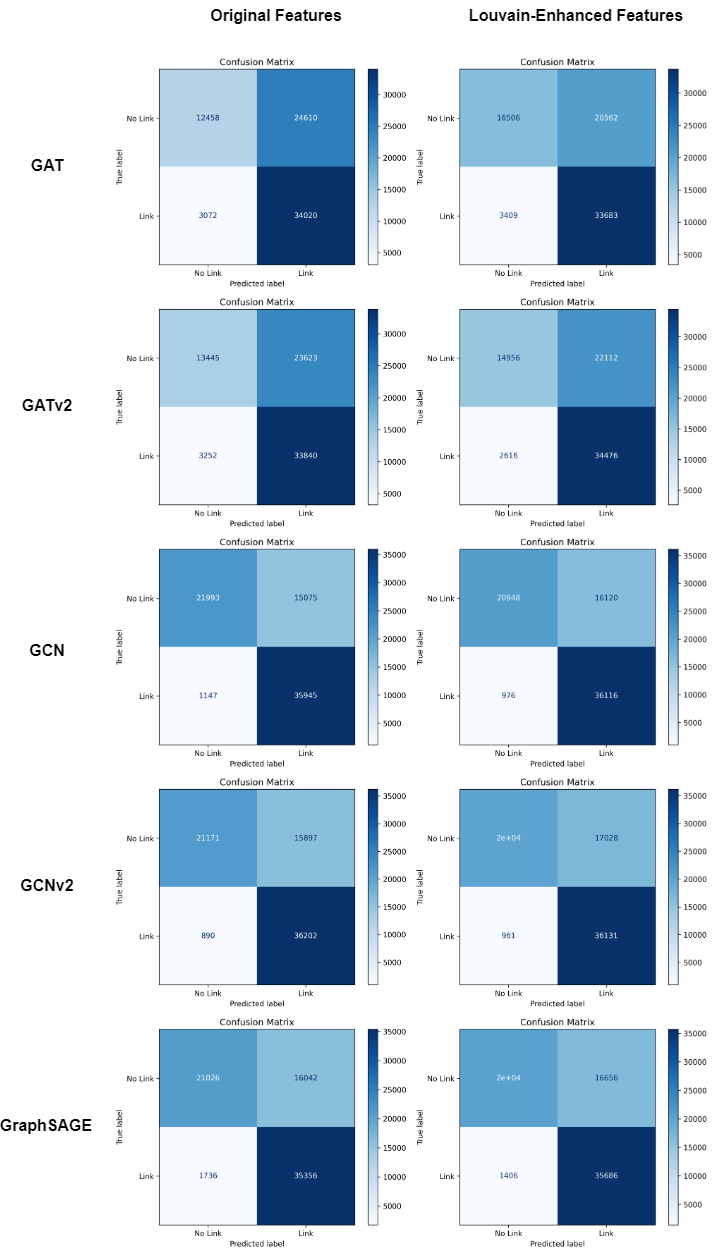}
\caption{Confusion matrices of different models. \label{confusion_matrices}}
\end{figure}

\section[\appendixname~\thesection]{Experimental Setup and Reproducibility}
In the interest of reproducibility and to support other researchers in building upon our findings, we hereby detail the computational resources and frameworks utilized in our experiments. Our models were implemented using PyTorch \cite{pytorch2019}, a widely-adopted machine learning framework known for its flexibility and efficiency in research prototyping and production deployment. All experiments were executed on a system outfitted with a single RTX 4060 GPU, ensuring a high degree of computational performance. 
% All appendix sections must be cited in the main text. In the appendices, Figures, Tables, etc. should be labeled, starting with ``A''---e.g., Figure A1, Figure A2, etc.

%%%%%%%%%%%%%%%%%%%%%%%%%%%%%%%%%%%%%%%%%%
\begin{adjustwidth}{-\extralength}{0cm}

\reftitle{References}

\end{adjustwidth}
\end{document}